\documentclass[aps,prl,reprint,superscriptaddress,longbibliography,amsfonts,amssymb,amsmath,showpacs]{revtex4-1}

\usepackage{graphicx}
\usepackage{hyperref}
\usepackage{color}
\newcommand{\blue}[1]{#1}

\newcommand*{\debye}{\lambda_{\text{D}}}
\newcommand{\gammap}{\dot{\gamma}}
\newcommand{\zetaslip}{\zeta_\text{slip}}
\newcommand{\zetano}{\zeta_\text{no-slip}}

\begin{document}

\title{Anomalous zeta potential in foam films}

\date{\today}

\author{Laurent Joly}
\email{laurent.joly@univ-lyon1.fr}
\author{Fran\c{c}ois Detcheverry}
\author{Anne-Laure Biance}
\affiliation{Institut Lumi\`ere Mati\`ere, UMR5306 Universit\'e Lyon 1-CNRS, Universit\'e de Lyon 69622 Villeurbanne, France}

\begin{abstract}
Electrokinetic effects offer a method of choice to control 
flows in micro and nanofluidic systems. 
While a rather clear picture of these phenomena exists now for the liquid-solid interfaces, 
the case of liquid-air interfaces remains largely unexplored. 
Here we investigate at the molecular level electrokinetic
transport in a liquid film covered with ionic surfactants. 
We find that the zeta potential, quantifying the amplitude of
electrokinetic effects, depends on the surfactant coverage in an
unexpected way. 
%We find that the dependence of the  zeta potential on the surfactant density is unexpected.  
First, it increases upon lowering surfactant coverage from saturation. 
Second, it does not vanish in the limit of low coverage, 
but instead approaches a finite value. 
This behavior is rationalized by taking into account the key
  role of interfacial hydrodynamics, together with an ion-binding mechanism. 
We point out implications of these results for the strongly debated measurements of zeta potential at free interfaces, 
and for electrokinetic transport in liquid foams. 
\end{abstract}


% 47.61.-k 	Micro- and nano- scale flow phenomena
% 47.57.jd 	Electrokinetic effects
% 83.50.Lh 	Slip boundary effects (interfacial and free surface flows)
% 47.11.Mn 	Molecular dynamics methods 

% 47.15.gm 	Thin film flows
% 47.55.N- 	Interfacial flows
% 47.55.dk 	Surfactant effects

\maketitle

%%%% introduction %%%
Electrokinetic (EK) phenomena take place in the vicinity of interfaces, 
where the presence of ionized groups results in a locally charged layer in the liquid. 
This electric double layer (EDL) can be set in motion by an electric field, 
eventually inducing through viscosity a flow known as electro-osmosis. 
Such EK effects, which also include streaming current and potential, 
are not only relevant in the biological realm~\cite{Daiguji:2010}, 
where charged or polar lipid bilayers are ubiquitous, 
but have also gained in the last decade major technological importance. 
Electric driving of liquids in micro/nano-channels has indeed become the method of
choice in many fluidic applications such as colloid or macromolecule
separation, or miniaturized energy conversion 
devices~\cite{Squires2005,Schoch2008,Bocquet2010}. 

As a coupling effect between electrostatics and hydrodynamics, 
EK effects, which are quantified by the zeta potential~\cite{Hunter2001}, 
depend not only on the electrostatic potential at the interface, 
but also on the boundary condition that applies there for the flow, 
possibly involving some slip~\cite{Bocquet2007}. 
Implications have been examined  theoretically~\cite{Muller1986, Joly2004}, 
as well as characterized experimentally~\cite{Bouzigues2008a,Audry:2010}
for the liquid-solid interface. 
In contrast, the effect of the hydrodynamic boundary condition in  the case of liquid-gas interfaces remains largely
unexplored. 
Yet, they depart in two important ways from their solid counterpart.
%First, their surface charge is often very high. 
First, whereas at solid walls a no-slip boundary condition usually applies,  
friction with the gas is very low, thus allowing for large slip. 
Secondly, charges are not fixed to a wall, but carried by species  such as surfactants, which are mobile. 
Both differences point to the importance of a complete
characterization of EK phenomena in those systems, widely
  encountered in industrial processes such as water purification
  through electrically driven bubbles~\cite{Burns1997}, mineral
  flotation and foam fractionation. %~\cite{Lemlich1968}. 
These potentially new effects could also be exploited to 
control bubble flow in liquid-filled microchannels and 
to design new self-assembled materials such as foams stable
against drainage~\cite{Bonhomme2013}, which is nowadays a subject of
active study~\cite{Miralles2014, Chevallier2013}. 
More fundamentally, as the sign and
  magnitude of the surface potential at an air (or oil)-water
  interface remains strongly debated~\cite{Vacha2007, Roger2012}, 
%More fundamentally, as a strong debate remains on the sign and
%  magnitude of the surface potential at an air (or oil)-water
%  interface~\cite{Vacha2007, Roger2012}, 
a careful analysis of the 
  relationship between zeta potential measurements often carried on
  ~\cite{Creux2009,Takahashi2005} and exact charge borne by these fluid
  interfaces must be performed. 

\blue{In order to get a better insight into these questions, this work
  investigates} 
%Therefore, in this work we investigate 
EK effects in foam films, 
where the surface charge is carried by ionic surfactants.  
While the distribution of such mobile surfactants will itself depend on the flow~\cite{Langevin2014}, 
and may vary according to the specific experimental setup considered, 
here we focus on the basic feature that is common to all situations: 
the relative motion between the surfactants and the liquid. 
Using molecular dynamics simulations of films of aqueous electrolytes coated with 
a typical surfactant, we characterize the zeta potential and find 
a dependence on surfactant coverage very different from that expected at a liquid-solid interface. 
We show that this behavior can be rationalized on the basis of simple arguments 
that account for the specificity of the liquid-air interface.

%%%% Model, systems and method %%%%

%
%%%%%%%%%%%%%%%%%%%%%%%%%%%%%%%%%%%%%%%%%%%%%%%%%%%%%%%%%%%%%%%%%%%%%%%%
\begin{figure}
\centering\includegraphics[width=\linewidth]{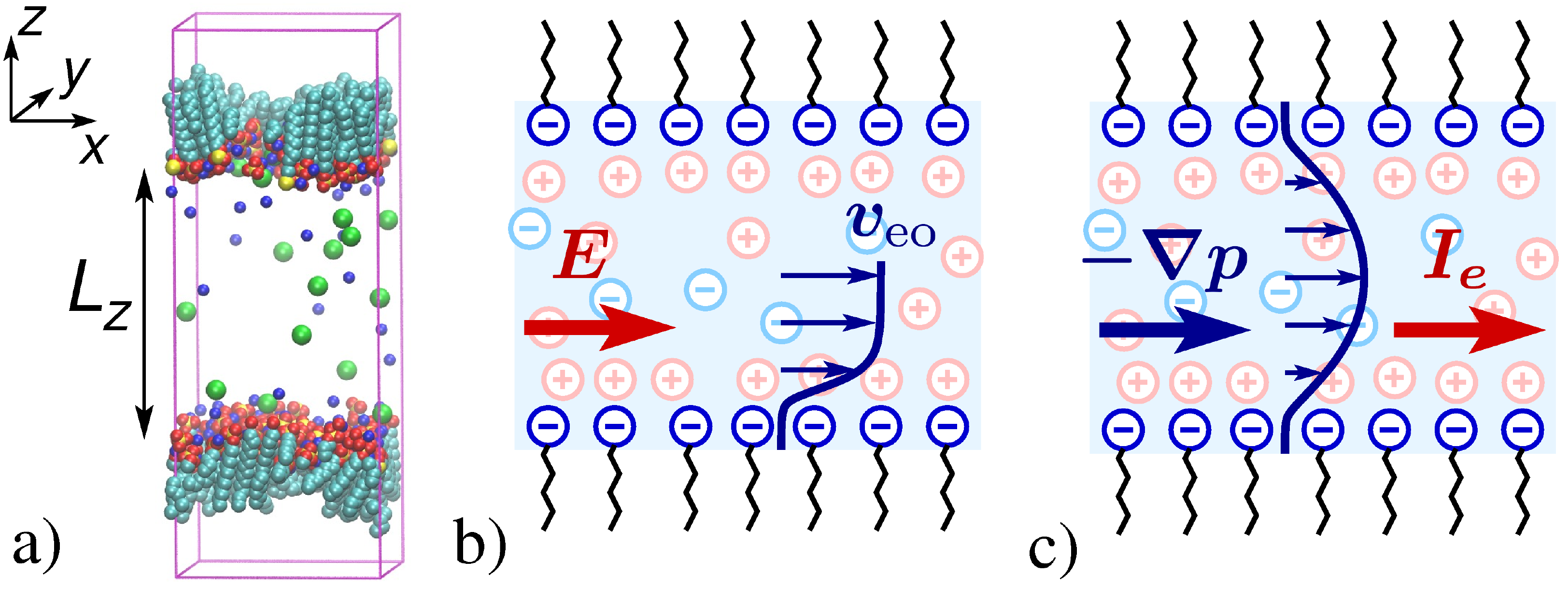}
\caption{a) Snapshot of a typical system ($\debye = 0.57$\,nm,
  $c=3.0$\,nm$^{-2}$); water molecules are not represented. 
b-c) Sketches of electro-osmosis (EO) and streaming current
  (SC) numerical experiments in foam films. } 
\label{fig:schema}
\end{figure}
%%%%%%%%%%%%%%%%%%%%%%%%%%%%%%%%%%%%%%%%%%%%%%%%%%%%%%%%%%%%%%%%%%%%%%%%
%

We considered water+salt (sodium chloride) films coated with sodium dodecyl
sulfate (SDS) surfactants, see Fig. \ref{fig:schema}a. 
Periodic boundary conditions were used 
in the film plane, with box dimensions $L_x = L_y = 4.6$\,nm. 
Water molecules, sodium ions and dodecyl
sulfate surfactants were modeled
following Bresme and Faraudo \cite{Bresme2004,Bresme2006}. In
particular, the SPC/E (extended simple point charge) model of water was
used, for its good dielectric and hydrodynamic representativity, at reasonably
low computational cost. An additional ingredient with regard to
Refs.~\cite{Bresme2004,Bresme2006} concerns chloride ions, which were modeled consistently with
sodium ions, using the parameters of Dang~\cite{Dang1995}.
Two salt concentrations have been considered, $\rho_s = 0.26$ and
$0.068$\,M, with corresponding Debye lengths (the width of the EDL) $\debye = 0.57$ and
$1.1$\,nm. The height of the films, along the $z$ direction, was fixed
to $L_z = 10\lambda_D$, in order to ensure no EDL overlap. 
For each salt concentration, the surface
density of surfactants $c$ was varied 
% by changing the number of surfactants in the unit cell, from 64 on each surface to 1, with a corresponding
% surface density $c$ ranging 
from $0.047$ to $3.0$\,nm$^{-2}$, with a corresponding 
surface charge $\Sigma$ ranging from $-7.6$ to $-480$\,mC/m$^2$. 
% Overall the number of atoms in the simulations ranged from 13'000 to
% 30'000. 
As a comparison, surface densities up to 2.2\,nm$^{-2}$ have been
measured experimentally in the absence of salt~\cite{Bergeron:1997}.
% However, on the timescale of simulation, exchange of surfactants with the liquid bulk is prohibited.
The simulations were performed using LAMMPS \cite{lammps}. Simulation
details can be found in supplemental material (SM). 

Two types of numerical experiments have been performed on these systems
(see Fig.~\ref{fig:schema}b-c): 
streaming current (SC) and electro-osmosis (EO). 
In the former configuration, a Poiseuille flow is induced in the $x$ direction, 
and the resulting electric current is measured. 
To induce the flow, a gravitylike force, adding up to $F$,  was applied to the liquid atoms, 
and a counterforce adding up to $-F$ was applied to the surfactant atoms. 
%Following our discussion in the introduction, 
%The total current $I_e$  due to ions and surfactants was measured, 
%a procedure equivalent to measuring the ionic current in the
%surfactants' reference frame, 
The ionic current $I_e$ was then measured in the
surfactants' reference frame, 
and the zeta potential computed from the standard formula:
$I_e/\mathcal{A} =  (\varepsilon \zeta / \eta) (-\nabla p)$. 
Here
$\mathcal{A} = L_y L_z$ is the film cross-section, 
$\varepsilon$ and
$\eta$ are the permittivity and dynamic viscosity of the liquid, 
and
$-\nabla p = F/(L_x L_y L_z)$ is the force applied to the liquid per unit volume. 
The viscosity was computed in the same simulations from the 
curvature of the Poiseuille velocity profile. %, depending slightly on
% $\debye$ and $c$ between $0.64$ and $0.69$\,mPa\,s. 
As regards the dielectric constant, we used the tabulated value for bulk SPC/E water at 300\,K, 
$\varepsilon_\text{r} = 70$~\cite{Reddy1989,Bonthuis2013}. 
EO numerical experiments were also performed, applying an electric field in
the $x$ direction, and measuring the resulting EO flow. 
The applied electric field $E_x$ ranged from $0.05$ to
$0.2$\,V\,nm$^{-1}$  
depending on the surfactant coverage, to ensure that the system
response remained linear with the forcing.
We considered the relative motion between the liquid and the surfactant layers to
compute the EO velocity $v_{eo}$ in the middle of the
film and obtained  the zeta potential from  
$v_{eo}/E_x = \varepsilon \zeta / \eta$.  
For each situation, we ran three independent simulations from distinct 
initial configurations, in order to reduce statistical
uncertainties.

%%%% Results %%%%

%
%%%%%%%%%%%%%%%%%%%%%%%%%%%%%%%%%%%%%%%%%%%%%%%%%%%%%%%%%%%%%%%%%%%%%%%%
\begin{figure}
\centering\includegraphics[width=0.9\linewidth]{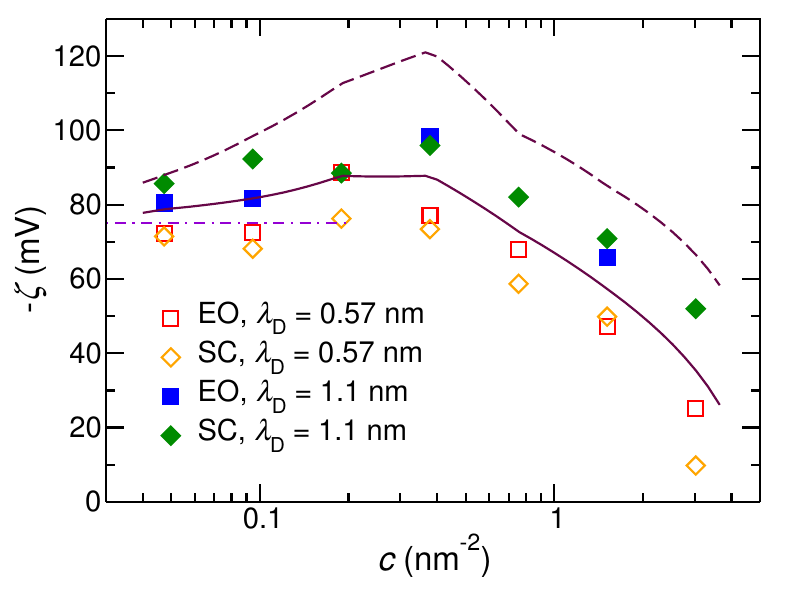}
\caption{ Zeta potential as a function of  surfactant coverage $c$,
  measured from EO and SC simulations, for two
   Debye lengths $\debye$. 
  The solid and dashed lines are predictions from Eq.~\eqref{eq:zeta},
  for $\debye = 0.57$ and $1.1$\,nm respectively. 
  The dot-dashed line is the dilute limit given by Eq.~\eqref{eq:zetadilute}.} 
\label{fig:zeta}
\end{figure}
%%%%%%%%%%%%%%%%%%%%%%%%%%%%%%%%%%%%%%%%%%%%%%%%%%%%%%%%%%%%%%%%%%%%%%%%
%

Figure~\ref{fig:zeta} presents the evolution of the zeta potential as a
function of the surfactant coverage, obtained for two
Debye lengths and from both EO and SC 
measurements. 
Within uncertainties, data from the two approaches match quantitatively,
as required by Onsager's reciprocal relations~\cite{Onsager1931a,Onsager1931}.
The results shown in Fig.~\ref{fig:zeta} display a number of
striking features. 
First, at high surfactant coverage,  the
zeta potential is much smaller than what could have been expected from
the large surface charge at stake. 
Then, when the surfactant concentration decreases, the zeta potential \emph{increases}. 
Finally, the zeta potential reaches a constant value in the limit of vanishing coverage. 

%
%%%%%%%%%%%%%%%%%%%%%%%%%%%%%%%%%%%%%%%%%%%%%%%%%%%%%%%%%%%%%%%%%%%%%%%%
\begin{figure}
\centering\includegraphics[width=\linewidth]{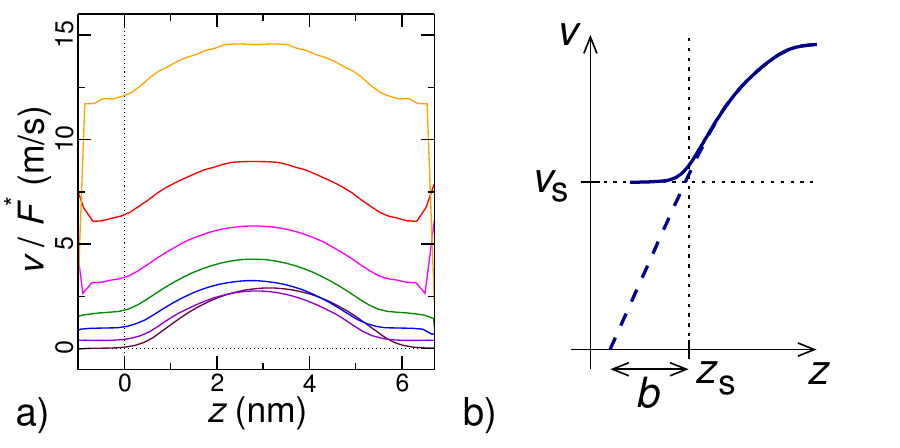}
\centering\includegraphics[width=\linewidth]{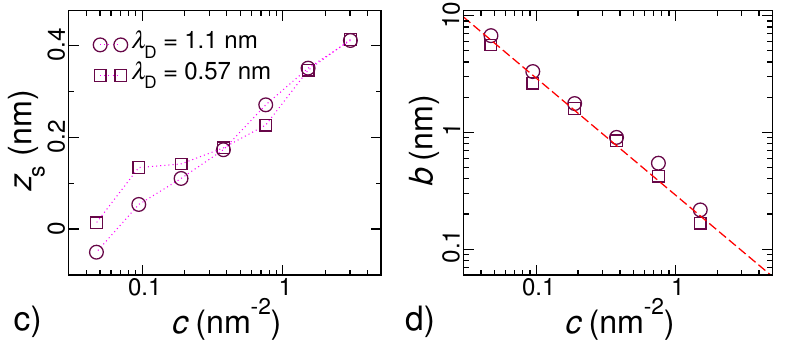}
\caption{Top: a) Poiseuille velocity profiles in the SC 
  configuration ($\debye =
  0.57$\,nm), for increasing surfactant coverage $c$ from top to bottom
  The velocity is measured in the surfactant 
  reference frame, and divided by the normalized applied force $F^*
  = F / F(c=0.047\,\text{nm}^{-2})$ for
  comparison purposes. The origin of the $z$ axis is taken at the average position of
  the surfactant  sulfur atoms. b) Cartoon illustrating the characterization of the
  hydrodynamic boundary condition. Bottom: Shear plane position $z_s$ (c) and slip length $b$ (d) as a function of the surfactant
  coverage, for two Debye lengths $\debye$. The slip length is fitted
  using Eq.~\eqref{eq:b} (dashed line).} 
\label{fig:pois}
\end{figure}
%%%%%%%%%%%%%%%%%%%%%%%%%%%%%%%%%%%%%%%%%%%%%%%%%%%%%%%%%%%%%%%%%%%%%%%%
%
To understand those results, we start by focusing on the hydrodynamic boundary condition at the interface, 
and  quantify the relative motion between the liquid and the surfactant layer. 
The latter is revealed more clearly in the SC case,
{\it i.e.}, when a Poiseuille flow is induced in the system. 
In Fig.~\ref{fig:pois}a we plot the liquid velocity profiles in the
reference frame of the surfactant layers 
for various surfactant coverages.  % (and for two Debye lengths $\debye= 5.7$ and $11$\,\AA). 
In the central part of the film, one can observe a characteristic parabolic
profile. However the liquid velocity does not vanish at the level of the
surfactant layers, as it would at most solid surfaces, where a no-slip boundary condition applies. 
Instead the velocity profile displays a plateau that extends across the
surfactant-laden interface. 
Keeping in mind that we plotted the
velocity profiles in the reference frame of the surfactants, this
plateau value corresponds to a velocity jump between the liquid and
the surfactant layer.
This velocity jump, or slip velocity, will be denoted by $v_s$ in the following.  
One can also define the shear plane, where the extrapolated
parabolic profile reaches $v_s$, and whose position will be denoted by
$z_s$. 
Finally, the observed velocity jump
is usually discussed in terms of the so-called partial slip boundary
condition~\cite{Bocquet2007}, which relates the slip velocity $v_s$ to the shear rate 
$\gammap(z) = \partial_z v$ at the
shear plane (see Fig.~\ref{fig:pois}b): 
$v_s = b\,\dot{\gamma} (z_s)$, with $b$ the slip length. 

Figure~\ref{fig:pois}c-d sums up our measurements of the shear plane position $z_s$ and slip length $b$,
for both Debye lengths considered and all surfactant densities. 
As observed and discussed earlier by some of us~\cite{Joly2004,Joly2006a}, 
the hydrodynamic boundary condition does not depend significantly on the Debye length. 
On the other hand, it is strongly affected by the surfactant coverage. 
As shown in Fig.~\ref{fig:pois}d, the slip length decreases as $b \propto c^{-1}$, 
a behavior that can be  rationalized with a simple picture~\cite{Ehlinger2013}.  
If the fluid moves with velocity $v_s$ with respect to the surfactant heads, 
the total friction force per unit area is $\mathcal{F} = \alpha \eta R v_s \, c$, 
where $R$ is a characteristic size, and $\alpha$ a dimensionless, geometric factor. 
Since by definition $\mathcal{F}=\eta/b \times v_s$ \cite{Bocquet2007}, one gets 
$b=1/(\alpha R c)$. % (see also Ref. \cite{Ehlinger2013} where this scaling
% law was obtained in a similar system). 
While this argument \textit{a priori} holds only in the dilute limit, 
when the contributions from each surfactant can be added, 
it describes the simulation data almost up to the highest surface coverage considered, 
i.e. close to saturation. 
For definiteness, the surfactant heads are now idealized as half-sphere, for which $\alpha=3 \pi$, 
giving
\begin{equation}
 b = \frac{1}{3 \pi \, R \,c},
\label{eq:b}
\end{equation}
where the hydrodynamics radius obtained by fitting the numerical results is $R=0.364$\,nm, 
the correct order of magnitude expected from the head dimensions. 
As regards the shear plane position  (Fig.~\ref{fig:pois}c), 
a monotonic shift toward the film interior is observed upon increasing the surfactant density. 
This trend can be qualitatively understood as detailed in the SM, 
and plays a significant role in the zeta potential, as shown below.

%
%%%%%%%%%%%%%%%%%%%%%%%%%%%%%%%%%%%%%%%%%%%%%%%%%%%%%%%%%%%%%%%%%%%%%%%%
\begin{figure}
\centering\includegraphics[width=\linewidth]{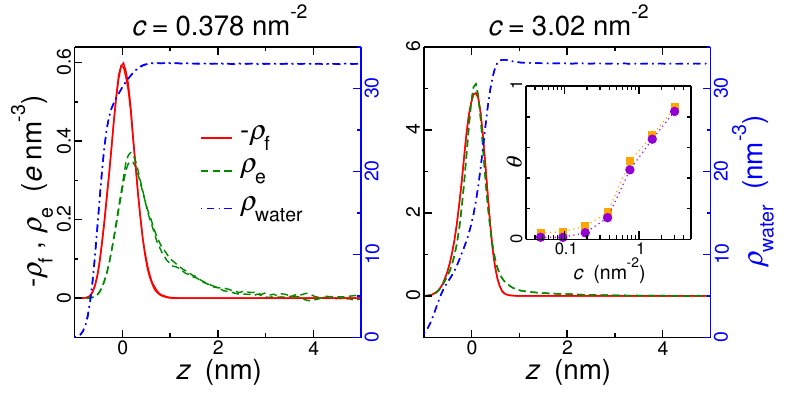}
\caption{%Top: 
 Typical charge density profiles of surfactants and
  ions, for $\debye =
  1.1$\,nm and two surfactant coverages $c$. The solid red lines
  represent the absolute surfactant charge $|\rho_f|=-\rho_f$. 
  The dashed green lines represent the ionic charge $\rho_e$. 
  The dash-dotted blue lines represent the water density profile. 
%  Bottom: corresponding electric potential profiles (see text for details). 
  Each data set is plotted for both the SC and
  EO simulations. The origin of the $z$ axis is taken at the average position of
  the surfactant sulfur atoms. 
  Inset: Fraction of bound ions $\theta$ versus surfactant coverage $c$, for $\debye$=1.1 and 0.57\,nm 
  (violet circles and orange squares respectively). } 
\label{fig:chargeprofiles}
\end{figure}
%%%%%%%%%%%%%%%%%%%%%%%%%%%%%%%%%%%%%%%%%%%%%%%%%%%%%%%%%%%%%%%%%%%%%%%%%%%%
%
We now turn to the ion distribution in the EDL. 
Figure~\ref{fig:chargeprofiles} displays typical charge and water density profiles for
low and high surfactant coverage, 
and shows that the surfactant-laden interface differs from a standard solid surface
in several ways. 
In contrast to the liquid-solid case,  
where charges are fixed to a wall of given geometry, 
here the surfactant heads carrying the charge can fluctuate in position with respect to the liquid-air interface, 
and the entire interface can deform due to capillary waves~\cite{jcp_125-014702-2006}. 
Both effects contribute to the widening of the charge distribution~\footnote{
Disentangling the two contributions would require a dedicated study and using the sophisticated 
methods developed recently to identify the intrinsic surface, see for
instance Ref.~\cite{pccp_10-4704-2008}.} observed in Fig.~\ref{fig:chargeprofiles}.  
Another striking feature is the nearly complete overlap 
between the ionic and surfactant charge distributions at high coverage. 
Analysis of numerical molecular configurations (see SM) shows that in
that case most ions are bound to the surfactant heads. 
The fraction $\theta$ of such bound ions, shown in the inset, vanishes at low coverage 
but approaches unity close to saturation. 
This ion binding can be described phenomenologically, as detailed in SM. 

Focusing now on the zeta potential, we extend previous approaches~\cite{Joly2004,Huang2007a,Huang2008} 
to the case of surfactant-laden interfaces. 
We consider the SC situation for simplicity 
but the results are directly transferable to EO, 
according to Onsager's reciprocal relations~\cite{Onsager1931a,Onsager1931}. 
The streaming current $I_e$ in the foam film is twice the current
originating at one surface, which can be written as the
integral over the interface of the electric current density:
$I_e = 2 L_y \int \rho_e(z) v(z) \mathrm{d}z$, 
where $L_y$ is the width of the interface, $\rho_e (z)$ the ionic charge
density and $v(z)$ the liquid velocity. 
The velocity profile is approximated as 
$ v(z) =  v_s +  \gammap (z-z_s) H(z-z_s)$, 
where $H$ is the Heaviside function, $v_s=\gammap b$, and 
$\gammap$ is the shear rate at $z_s$. 
\blue{This profile is the superposition of a plug flow at constant velocity $v_s$, 
and a ``no-slip'' part that neglects the curvature of the
velocity profile at the scale of the EDL. 
Accordingly, the zeta potential can be decomposed into slip and no-slip
contributions, $\zeta = \zetaslip + \zetano$. The slip contribution
corresponds to the plug flow of the whole ionic charge $\int \rho_e \mathrm{d}z = -\Sigma$ at
velocity $v_s$, and writes $\zetaslip = \Sigma
    b / \varepsilon$, independently of the ionic charge
distribution. To evaluate the no-slip contribution, as a first approximation, we neglect the spatial distribution of surfactant and take 
$\rho_f(z) = \Sigma \delta(z)$. 
Following traditional approaches~\cite{Hunter2001,Bocquet2010} yields
$\zetano =  V(z_s)$, where $V$ is the electrostatic potential. The
total zeta potential then reads: 
\begin{equation}
  \zeta = \zetaslip + \zetano = \frac{\Sigma
    b}{\varepsilon} + V(z_s).  
\label{eq:zeta}
\end{equation}
}
%where $\zetaslip$ denotes the contribution from the plug flow, 
%and $V$ is the electrostatic potential. 

\blue{In the limit of low surfactant coverage, 
the no-slip contribution $\zetano$ becomes negligible. 
Indeed, as we checked  from numerical results by charge density integration, 
it approaches the Debye-H\"{u}ckel result $\Sigma \lambda_D / \varepsilon$, 
as soon as  $\Sigma \lesssim 30$\,mC/m$^2$. The no-slip contribution
being proportional to the surface charge, it vanishes in the dilute
limit. 
On the other hand,  combining Eqs.~\eqref{eq:b} and~\eqref{eq:zeta}, 
and taking $\Sigma = - e c$, since ion-binding is negligible at low coverage, 
one gets 
\begin{equation}
\zeta (c \rightarrow 0) = -\frac{e}{3\pi R \epsilon}. 
\label{eq:zetadilute}
\end{equation}
Interestingly, the surface charge and slip length dependence on the
surfactant coverage in the slip contribution compensate exactly in the
dilute limit,   
% With the surface charge and the no-slip contribution  vanishing, 
% the total zeta potential therefore identify with the slip
% contribution, 
and the zeta potential reaches a finite value for vanishing surfactant
coverage. 
Even though saturation of slip length can result in non-monotonic variation
of the zeta potential~\cite{Joly2006a,Jing2013,Chakraborty2013}, 
a finite value  in the limit of vanishing surface charge is 
unknown for liquid-solid interface.} 
Furthermore, a numerical estimate of Eq.~\eqref{eq:zetadilute} yields a value of $-75$\,mV, 
in  good agreement with numerical results, see Fig.~\ref{fig:zeta}.   
Importantly, this  suggests that even very few impurities on a bare interface 
can generate a non-negligible zeta potential, 
whose magnitude depends on impurity size and charge. 
This effect might play an important
role in the understanding of EK measurements of surface charge 
near free interfaces~\cite{Roger2012}.

Finally, we use Eq.~\eqref{eq:zeta} to estimate the zeta potential 
over the whole range of surfactant coverage. 
In doing so, we assume that bound ions completely cancel the surfactant charge, 
leading to an apparent charge $\Sigma=-ec(1-\theta)$. 
For the potential $V(z)$, we take the exact solution to the Poisson-Boltzmann equation for a single wall. 
Taking Eq.~\eqref{eq:b} for the slip length, 
and the simulation results for $z_s$ and $\theta$, 
yields the theoretical curves shown in Fig.~\ref{fig:zeta}. 
While they consistently overestimate the simulation results, 
they capture the main trend as a function of coverage and Debye length, 
with a collapse at high~$c$ induced by ion binding. 
In view of the crudeness of the model (see SM for an improved but somewhat {\it ad hoc} description), 
the agreement is reasonable. This suggests 
that to recover the unusual dependence of the zeta potential, 
the three main ingredients are:
the slip length dependence $b \propto c^{-1}$, 
the shift in shear plane position $z_s$ 
and the ion binding. 
% Note finally that %in contrast with the dilute regime, 
% %the zeta potential at intermediate and high coverage may exhibit 
% %qualitatively different behavior according to the surfactant considered. 
% depending on the $\theta(c)$ dependence, which is set also by the nature of surfactants (see SM),
% $\zeta(c)$ may be monotonously decreasing, or exhibit a peak at intermediate coverage. 

As a conclusion, we have characterized at the molecular level  
electrokinetic effects at a foam film interface, and its dependence on surfactant coverage. 
We find a nontrivial and nonconventional dependence, the zeta potential
tending to decrease upon increasing the  surface charge. 
Strikingly, in the dilute limit, the slippage contribution 
compensates  exactly  for the decrease in surface charge, 
resulting in a saturation value of the zeta potential
around $75$\,mV in our case. 
This value is significant as experimental values for zeta potential 
typically fall in the range 0-150\,mV. 
Because they point out the  key role of
impurities even at very low density, 
our findings are relevant for the understanding of surface potential measurements 
on free interfaces  \cite{Marinova1996, Roger2012}, most notably water \cite{Creux2009,Takahashi2005},
which remain highly debated \cite{Vacha2007}. 
Overall, this study is a first step toward a complete understanding of
electrokinetics near surfactant-laden interfaces. 
Having characterized locally the relative motion between liquid and surfactants, 
one can now address the situation where the global surfactant distribution is inhomogeneous, 
and induces Marangoni flow. 
Electrokinetics with surfactants as mobile charge carriers may induce a variety of effects, 
all at play in a liquid foam.

\begin{acknowledgments}
%{\it Acknowledgments}
We thank Lyd\'{e}ric Bocquet and
Christophe Ybert 
for fruitful discussions throughout the completion of this work. 
This study was supported by the French Agence Nationale de la Recherche (ANR) through the 
E-FOAM Project. 
We are grateful for the computing resources of JADE (CINES, French
National HPC) obtained through Project No. c20132a7167, and of the PSMN
(P\^ole Scientifique de Mod\'elisation Num\'erique) Computing Center
of ENS de Lyon. 
\end{acknowledgments}

%\bibliography{bibliolaurent,bibliolaurent1,biblioadd1}

\end{document}